\documentstyle[12pt]{article}
\newcommand{\Tr}{\hbox{Tr}}
\newcommand{\Str}{\hbox{Str}}
\newcommand{\del}{\partial}
\newcommand{\dis}{\displaystyle}
\newcommand{\scr}{\scriptsize}

\newcommand{\1}{\bf\mbox{1}}
\newcommand{\2}{\bf\mbox{i}}
\newcommand{\3}{\bf\mbox{j}}
\newcommand{\4}{\bf\mbox{k}}

\newcommand{\F}{\cal{F}}
\newcommand{\A}{\cal{A}}
\newcommand{\V}{\cal{V}}
\newcommand{\E}{\cal{E}}

\newcommand{\Q}{Q_{\hbox{\scriptsize B}}}
\newcommand{\Qt}{\widetilde{Q}}

\newcommand{\C}{\widetilde{C}}
\newcommand{\etab}{\overline{\eta}}

\newcommand{\delg}{\delta_{\hbox{\tiny gauge}}}
\newcommand{\st}{\widetilde{s}}

\renewcommand{\theequation}{\arabic {section}.\arabic{equation}}

\makeatletter
\def\eqnarray{%
 \stepcounter{equation}%
 \let\@currentlabel=\theequation
 \global\@eqnswtrue
 \global\@eqcnt\z@
 \tabskip\@centering
 \let\\=\@eqncr
 $$\halign to \displaywidth\bgroup\@eqnsel\hskip\@centering
 $\displaystyle\tabskip\z@{##}$&\global\@eqcnt\@ne
 \hfil$\displaystyle{{}##{}}$\hfil
 &\global\@eqcnt\tw@$\displaystyle\tabskip\z@{##}$\hfil
 \tabskip\@centering&\llap{##}\tabskip\z@\cr}
\makeatother

\makeatletter
\def\@arrayacol{\edef\@preamble{\@preamble \hskip .5\arraycolsep}}
\def\array{\let\@acol\@arrayacol \let\@classz\@arrayclassz
\let\@classiv\@arrayclassiv \let\\\@arraycr\def\@halignto{}\@tabarray}
\makeatother

\renewcommand{\arraystretch}{1.6}

\begin{document}

\setlength{\baselineskip}{7mm}
\begin{titlepage}
\begin{flushright}
EPHOU-99-008 \\ 
May, 1999
\end{flushright}
 
\vspace{2cm}

\begin{center} 
{\Large $N=2$ Supersymmetric Model with Dirac-K\"ahler Fermions}\\
{\Large from Generalized Gauge Theory in Two Dimensions}\\
\vspace{1cm}
{\bf{\sc Noboru Kawamoto and Takuya Tsukioka}}\\
{\it{Department of Physics, Hokkaido University}}\\
{\it{Sapporo, 060-0810, Japan}}\\
{kawamoto, tsukioka@particle.sci.hokudai.ac.jp}
\end{center}
\vspace{2cm}

\begin{abstract}
We investigate the generalized gauge theory which has been proposed  
previously and show that in two dimensions the instanton gauge fixing 
of the generalized topological Yang-Mills action leads to 
a twisted $N=2$ supersymmetric action. 
We have found that the $R$-symmetry of $N=2$ supersymmetry 
can be identified with the flavour symmetry of Dirac-K\"ahler 
fermion formulation. 
Thus the procedure of twist allows topological ghost fields 
to be interpreted as the Dirac-K\"ahler matter fermions.  
\end{abstract}

\end{titlepage}

\newpage

\section{Introduction}

\setcounter{equation}{0}
\setcounter{footnote}{0}

In formulating unified theory it is the 
general consensus that the supersymmetry may play a crucial.
It is important to understand the origin of supersymmetry and 
the fermion and boson correspondence. 
There is an interesting example of topological field theory analysis 
by Witten\cite{w1} which suggests the possible origin of $N=2$ supersymmetry 
and the generation of fermionic fields from ghost via twisting 
procedure. 
Later it has been pointed out that this theory 
can be derived from the ``partially'' BRST gauge-fixed action 
of the topological Yang-Mills action 
with instanton gauge fixing~\cite{lp, bs}.   
This subject has been intensively investigated~\cite{bbrt}, 
particularly in connection with supersymmetric field
theories~\cite{brt}.
In this paper we claim that the topological twist generating the 
matter fermions from ghosts is essentially related to the 
Dirac-K\"ahler fermion formulation. 

In 1960's K\"ahler~\cite{k} has shown 
that Dirac equation is constructed from inhomogeneous differential forms 
which are called Dirac-K\"ahler fields~\cite{g}.
Moreover Dirac-K\"ahler fermion is a curved spacetime version of 
Kogut-Susskind fermion~\cite{kss} 
or staggered fermion~\cite{ks} 
and thus a natural framework 
of the lattice fermion formulation~\cite{bj}. 

About ten years ago one of the authors (N.K.) and Watabiki 
proposed a generalization of the ordinary three-dimensional 
Chern-Simons theory into arbitrary dimensions 
by introducing all the degrees of differential forms 
as gauge fields and parameters together with quaternion 
structure~\cite{kw1}. 
Later the quantization of the even-dimensional version 
of the generalized Chern-Simons actions has been completed 
by Batalin-Vilkovisky formulation~\cite{kostu}. 
This formulation can be, however, generalized to topological 
Yang-Mills and ordinary Yang-Mills actions. 

Since the generalized gauge theory is formulated by differential forms 
it has close connection with the Dirac-K\"ahler fermion formulation.
We believe that the generalized gauge theory may play a crucial 
role in formulating the unified model including quantum gravity 
on the simplicial lattice manifold~\cite{kw2}.

In this paper we investigate the generalized topological
Yang-Mills theory from the topological field theory point of view.  
An enlarged algebraic structure of BRST transformations 
\'a la Baulieu-Singer~\cite{bs} is  
naturally constructed in a unified way 
by the generalized gauge theory. 
As the simplest example towards more realistic case, 
we quantize the two-dimensional version of the generalized 
topological Yang-Mills action and show that  
the ``partially'' gauge-fixed action with instanton gauge fixing 
leads to a twisted $N=2$ 
supersymmetric abelian Higgs action~\cite{dvf, st} 
without symmetry breaking potential term. 
It is interesting to recognize that our instanton relations coincide 
with dimensionally reduced Seiberg-Witten equations~\cite{sw} 
from four into two dimensions~\cite{ns}.
We point out that the fermionic ghost fields 
can be interpreted as Dirac-K\"ahler fermion fields and 
thus the twisting procedure is nothing but the Dirac-K\"ahler 
fermion formulation.   

This paper is organized as follows.  
In section 2 we summarize the generalized gauge theory 
in arbitrary dimensions. 
In section 3 we analyze the generalized two-dimensional 
topological Yang-Mills theory as a topological field theory.
In section 4 we explicitly verify the twisted $N=2$ supersymmetric 
algebra for the gauge-fixed action.  
In section 5 we explain the twisting mechanism via  
Dirac-K\"ahler formulation.  
Conclusions and discussions are given in the final section.

\section{Generalized gauge theories in arbitrary dimensions}

\renewcommand{\theequation}
{\arabic{section}.\arabic{equation}}
\setcounter{equation}{0} 

In this section we summarize the formulation of the generalized gauge
theory with an emphasis on their algebraic structures.  

The essential point of the generalization is 
to extend a one-form gauge field and zero-form gauge parameter 
to a quaternion valued generalized gauge field 
and gauge parameter which contain forms of all possible degrees. 
 
In the most general form, a generalized gauge field ${\cal A}$ and a
gauge parameter ${\cal V}$ are defined by the following component form:
\begin{eqnarray}
{\A} &=& {\1}\psi + {\2}\hat{\psi}
            +{\3}A + {\4}\hat{A}, \\
{\V} &=& {\1}\hat{a} + {\2}a
            +{\3}\hat{\alpha} + {\4}\alpha,
\end{eqnarray}
where $(\psi, \alpha)$, $(\hat{\psi}$, $\hat{\alpha})$, $(A, a)$ and
$(\hat{A},\hat{a})$ are direct sums of fermionic odd forms, fermionic
even forms, bosonic odd forms and bosonic even forms, respectively, and
they take values on a gauge algebra.  
The symbols ${\1}$, ${\2}$, ${\3}$ and ${\4}$ satisfy the
following quaternion algebra: 
\begin{equation}
 \begin{array}{c}
 {\1}^2 = {\1}, \quad {\2}^2 = -{\1}, \quad 
 {\3}^2 = -{\1}, \quad {\4}^2 = -{\1}, \\
 {\2}{\3}=-{\3}{\2}={\4}, \quad
 {\3}{\4}=-{\4}{\3}={\2}, \quad
 {\4}{\2}=-{\2}{\4}={\3}. \\
 \end{array}
\end{equation}
The following graded Lie algebra can be adopted as a gauge algebra: 
\begin{eqnarray}
 {[T_a, T_b]} &=& f^c_{ab}T_c, \nonumber \\
 {[T_a, \Sigma_\beta]} &=& g^\gamma_{a \beta}\Sigma_\gamma, \\
 \{ \Sigma_\alpha, \Sigma_\beta \} &=&
h^c_{\alpha\beta}T_c, \nonumber 
\end{eqnarray}
where all the structure constants are subject to consistency conditions
which follow from the graded Jacobi identities.  
The components of the gauge field ${\A}$ and the gauge parameter ${\V}$ 
are particularly assigned as elements of the gauge algebra
\begin{equation}
 \begin{array}{rclcrclcrclcrcl}
 A &=& T^aA_a, &\quad& \hat{\psi} &=& T^a\hat{\psi}_a, &\quad&
 \psi &=& \Sigma^\alpha\psi_\alpha, &\quad& 
 \hat{A} &=& \Sigma^\alpha\hat{A}_\alpha, \\
 \hat{a} &=& T^a\hat{a}_a, &\quad& \alpha &=& T^a\alpha_a, &\quad&
 \hat{\alpha} &=& \Sigma^\alpha\hat{\alpha}_\alpha, &\quad& a &=&
 \Sigma^\alpha a_\alpha. \\
 \end{array}
\end{equation}
The component expansion of the same type as ${\A}$ and ${\V}$
are classified as elements of $\Lambda_-$-class and 
$\Lambda_+$-class, respectively.
These elements fulfill the $Z_2$-grading structure
\begin{equation}
 {[\lambda_+, \lambda_+]}\in \Lambda_+, \quad {[\lambda_+,
 \lambda_-]}\in \Lambda_-, \quad \{\lambda_+, \lambda_+\}\in \Lambda_+,
\end{equation}
where $\lambda_+\in\Lambda_+$ and $\lambda_-\in\Lambda_-$.  
In particular the exterior derivative belongs to $\Lambda_-$-class
\begin{equation}
 Q = {\3}d,
\end{equation}
and the following relations similar to the ordinary exterior derivative 
operator hold
\begin{equation}
 Q(\lambda_1\lambda_2)=(Q\lambda_1)\lambda_2
                       +(-)^{|\lambda_1|}\lambda_1(Q\lambda_2), \quad 
 Q^2 = 0, 
\end{equation}
where $|\lambda_1|=0$ for $\lambda_1\in\Lambda_+$ and 
$|\lambda_1|=1$ for $\lambda_1\in\Lambda_-$. 
To construct the generalized actions, 
the two types of traces for the gauge algebra 
should be introduced, 
\begin{equation}
 \Tr{[T^a, \cdots]}=0, \quad 
 \Tr{[\Sigma^\alpha, \cdots]}=0,
\end{equation}
\begin{equation}
 \Str{[T^a, \cdots]}=0, \quad 
 \Str\{\Sigma^\alpha, \cdots\}=0,
\end{equation}
where $(\cdots)$ in the commutators and the anticommutator denote a
product of the generators.  
These definitions of the traces are crucial
so that the generalized actions are invariant 
under the generalized gauge transformations.

We can then construct generalized actions in terms of these generalized 
quantities. 
The generalized Chern-Simons actions 
which have been previously proposed~\cite{kw1}  
are given on even- and odd-dimensional manifolds $M$,  
\begin{eqnarray}
 S_{\mbox{\scr{even}}} &=& \int_M\Tr_{\4}
                  \bigg( {\A}Q{\A}+\frac{2}{3}{\A}^3 \bigg), 
\label{eq:egcs} \\
 S_{\mbox{\scr{odd}}} &=& \int_M\Str_{\3}
                  \bigg( {\A}Q{\A}+\frac{2}{3}{\A}^3 \bigg),
\end{eqnarray}
where $\Tr_{\4}(\cdots)$ and
$\Str_{\3}(\cdots)$ are defined so as to pick up only
the coefficient of ${\4}$ and ${\3}$ from $(\cdots)$
and take the traces.  
We then need to pick up $d$-form terms
corresponding to $d$-dimensional manifolds $M$.  
These actions are invariant up to surface terms 
under the following generalized gauge transformation: 
\begin{equation}    
 \delta{\A}={[Q+{\A}, {\V}]},
\end{equation}
where ${\V}$ is the generalized gauge parameter.  
It should be noted that this symmetry is much larger 
than the usual gauge symmetry 
since the gauge parameter ${\V}$ contains many parameters of various
forms.

There is another suggestive topological nature due to 
the parallel construction to the standard gauge theory. 
In the generalized gauge theory it is possible to define 
generalized Chern character which is expected to have 
topological nature 
\begin{eqnarray} 
 \Str_{\1}({\F}^n)
 &=&\Str_{\1}(Q\Omega_{2n-1}), 
\label{eq:bepo} \\
 \Tr_{\2}({\F}^n)
 &=&\Tr_{\2}(Q\Omega_{2n-1}), 
\label{eq:bopo}  
\end{eqnarray}
where ${\F}$ is a generalized curvature 
\begin{equation}
 {\F}=Q{\A}+{\A}^2, 
\end{equation}
and $\Omega_{2n-1}$ are the ``generalized'' Chern-Simons form. 
Eqs. (\ref{eq:bepo}) and (\ref{eq:bopo}) are bosonic even form and  
bosonic odd form, respectively. 
Especially, for $n=2$ case in (\ref{eq:bepo}), 
we obtain a topological Yang-Mills type action related to 
a one dimension lower generalized Chern-Simons action  
on an even-dimensional manifold $M$,
\begin{equation}
 \int_M{\mbox{Str}}_{\1}{\F}^2
 =\int_M{\mbox{Str}}_{\1}\Bigg( Q\Big({\A}Q{\A}
 +\frac{2}{3}{\A}^3\Big)\Bigg), 
\label{eq:2cc}
\end{equation}
which has the same forms of the standard relation. 

\section{Generalized topological Yang-Mills theory in two dimensions}

\setcounter{equation}{0} 

In this section we analyze the two-dimensional version of 
the generalized topological Yang-Mills action. 
Our formulation of this section is the two-dimensional realization 
of the known four-dimensional scenario~\cite{w1, lp, bs} 
and can be extended to arbitrary dimensions.  

As we have already mentioned that the action we consider satisfies 
the following well known relation: 
\begin{equation}
 \int_M{\mbox{Str}}_{\1}{\F}_0^2
 =\int_M{\mbox{Str}}_{\1}\Bigg( Q\Big( {\A}_0Q{\A}_0
                                       +\frac{2}{3}{\A}_0^3 \Big) \Bigg),
\end{equation}
where ${\A}_0$ and ${\F}_0$ are the two-dimensional counter part 
of the classical gauge field and curvature. 
More explicitly they are given by 
\begin{eqnarray}
 {\A}_0 &=& {\3}\omega 
           +{\4}(\phi+B), 
 \qquad\in\Lambda_-, \\
\label{eq:A0}
 {\F}_{0}&=& Q{\A}_0 + {\A}_0^2 \nonumber \\
         &=& -{\1}\Big( d\omega + \omega^2 + \{\phi, B\} + \phi^2 \Big) 
            +{\2}\Big(d\phi+[\omega, \phi]\Big),  
\qquad\in\Lambda_+, 
\label{eq:F0}
\end{eqnarray}
where $\phi$, $\omega$ and $B$ are graded Lie algebra
valued zero-, one- and two-form gauge fields, respectively.
Due to the topological nature of the action, the action has so called 
shift symmetry. 
In other wards the action is invariant under the arbitrary deformation 
of the gauge field ${\A}_0$, which we denote ${\E}_0$.  
Thus the gauge transformation of the generalized topological 
Yang-Mills action has the following form: 
\begin{equation}
 \delta{\A}_{0} = [Q+{\A}_0, {\V}_0]+{\E}_0, 
\label{eq:gt}
\end{equation}
where ${\V}_0$ is the generalized gauge parameter 
\begin{equation}
 {\V}_0 = {\1}(v+b)+{\2}u, \qquad\in\Lambda_+,
\end{equation}
while ${\E}_0$ is a new gauge parameter of the shift symmetry 
and is given by 
\begin{equation}
 {\E}_0={\3}\xi_{(1)}+{\4}(\xi_{(0)}+\xi_{(2)}),
 \qquad\in\Lambda_-, 
\end{equation}
where the suffix $(n)$ with $n=0, 1, 2$ denotes the form degree. 
Hereafter we use the same notation to the form degree. 
The field strength is transformed 
under the gauge transformation (\ref{eq:gt}),
\begin{equation}
 \delta{\F}_0=[{\F}_0, {\V}_0]+\{ Q+{\A}_0, {\E}_0 \}.
\label{eq:ift}
\end{equation}
The first term is transformed covariantly,  
and the second term is inhomogeneous gauge transformation 
of ${\F}_0$ by the gauge parameter which belongs to $\Lambda_-$-class.

The topological shift symmetry of ${\cal E}_0$, 
however, can absorb the usual gauge transformation, 
so that this is a reducible system with the following 
obvious reducibility conditions:   
\begin{equation}
\begin{array}{rcl}
 {\V}_0 &=& {\V}_1, \\
 {\E}_0 &=& -[Q+{\A}_0, {\V}_1]. \\ 
\end{array}
\label{eq:reducibility}
\end{equation}
Correspondingly we need to introduce ghost fields with respect to 
the generalized gauge symmetry and the topological shift symmetry, 
and ghost for ghost fields with respect to 
the additional gauge symmetry of the gauge
parameter (\ref{eq:reducibility}). 

Although we can construct the nilpotent BRST algebra of the above reducible
system by the procedure of cohomological perturbation~\cite{bv},
we can treat it in an algebraically unified way by using 
the characteristic of the generalized gauge system. 
We redefine the generalized gauge field by introducing 
the generalized ghost fields $C_{(0)}$, $C_{(1)}$ and $C_{(2)}$   
\begin{equation}
 {\A} = {\1}C_{(1)}+{\2}(C_{(0)}+C_{(2)})+{\3}\omega+{\4}(\phi+B),
 \qquad\in\Lambda_-.
\label{eq:gfa}
\end{equation}
We need to introduce a generalized field which belongs another class 
of $\A$ to accomodate the topological ghost fields $\C_{(0)}$, $\C_{(1)}$ 
and $\C_{(2)}$  
and the ghost for ghost fields $\eta_{(0)}$, $\eta_{(1)}$ 
and $\eta_{(2)}$  
\begin{equation}
 {\cal{C}}={\1}(\eta_{(0)}+\eta_{(2)})+{\2}\eta_{(1)}
          +{\3}(\C_{(0)}+\C_{(2)})+{\4}\C_{(1)}, 
 \qquad\in\Lambda_+. 
\label{eq:gfc}
\end{equation}
Here ${\cal C}$ belongs $\Lambda_+$ and could be identified 
as a part of generalized curvature later. 

Furthermore we extend the concept of the differential operator 
by introducing the BRST operator $s$ as a fermionic
zero-form,\footnote{
The fermionic operator $s$ acts as a left derivative on fields 
in the same way as the operation of the exterior derivative $d$.
}
\begin{equation}
 {\cal{Q}} \equiv Q+\Q  = {\3}d+{\2}s, \qquad\in\Lambda_-.
\end{equation}
It should be noted 
that $s$ commutes with $d$, $i.e.$ $[d, s]=0$ and $s^2=0$.
This operator satisfies the nilpotency property 
due to the quaternion structures,
\begin{equation}
{\cal{Q}}^2 = 0.
\label{eq:np}
\end{equation}
The following graded Leibnitz rule acting on generalized gauge fields 
can be derived: 
\begin{equation}
 {\cal{Q}}(\lambda_1\lambda_2)
 =({\cal{Q}}\lambda_1)\lambda_2
    +(-)^{|\lambda_1|}\lambda_1({\cal{Q}}\lambda_2),
\label{eq:lr} 
\end{equation}
where $|\lambda_1|=0$ for $\lambda_1\in\Lambda_+$ and 
$|\lambda_1|=1$ for $\lambda_1\in\Lambda_-$. 

We can now construct BRST transformation 
algebraically in a unified way.
We define the generalized curvature by using the redefined gauge field 
\begin{eqnarray}
 {\F}&=&{\cal{Q}}{\A}+{\A}^2 \nonumber \\
     &=&{\F}_0 + {\cal{C}},  
\label{eq:gecu}
\end{eqnarray}
where the second relation is imposed to relate the BRST transformation 
with respect to classical and generalized ghost fields. 
${\cal{C}}=0$ corresponds to impose usual horizontal conditions.  
While the transformations with respect to the topological ghost 
and ghost for ghost fields are derived by 
the following Bianchi identity of the generalized field,  
\begin{equation}
 {\cal{Q}}{\F}+[{\A}, {\F}]=0.
\label{eq:geBi}
\end{equation}
The component wise expressions of BRST transformation can be read 
from (\ref{eq:gecu}) and (\ref{eq:geBi}):   
\begin{equation}
 \begin{array}{rcl}
 s\phi &=& -[C_{(0)}, \phi] - \C_{(0)}, \\
 s\omega &=& dC_{(0)}+[\omega, C_{(0)}] 
            -\{C_{(1)}, \phi\} + \C_{(1)}, \\
 sB &=& dC_{(1)}+\{\omega, C_{(1)}\} - [C_{(0)}, B ]
        -[C_{(2)}, \phi] - \C_{(2)}, \\
 sC_{(0)} &=& -C_{(0)}^2 - \eta_{(0)}, \\
 sC_{(1)} &=& -\{C_{0}, C_{(1)}\} + \eta_{(1)}, \\
 sC_{(2)} &=& C_{(1)}^2 - \{C_{(0)}, C_{(2)} \} - \eta_{(2)}, \\
 s\C_{(0)} &=& - \{C_{(0)}, \C_{(0)} \} - [\phi, \eta_{(0)} ], \\
 s\C_{(1)} &=& d\eta_{(0)}+[\omega, \eta_{(0)}] + [C_{(1)}, \C_{(0)}]
              -\{C_{(0)}, \C_{(1)}\} + \{\phi, \eta_{(1)}\}, \\
 s\C_{(2)} &=& d\eta_{(1)}+\{\omega, \eta_{(1)}\}-[C_{(1)}, \C_{(1)}]
              -\{C_{(0)}, \C_{(2)}\}-\{C_{(2)}, \C_{(0)}\} \\
           & &-[\phi, \eta_{(2)}] - [B, \eta_{(0)}], \\
 s\eta_{(0)} &=& - [C_{(0)}, \eta_{(0)}], \\
 s\eta_{(1)} &=& [C_{(1)}, \eta_{(0)}] - [C_{(0)}, \eta_{(1)}], \\
 s\eta_{(2)} &=& -[C_{(1)}, \eta_{(1)}] - [C_{(0)}, \eta_{(2)}] 
            - [C_{(2)}, \eta_{(0)}].  
 \end{array}
\label{eq:BRST11}
\end{equation}

These algebraic and geometric constructions 
of BRST transformation were emphasized 
by Baulieu-Singer~\cite{bs} 
for four-dimensional topological Yang-Mills model.
We here propose the natural extension of their approach 
in the framework of the generalized gauge theory.
Moreover we do not have to introduce the ghost number 
for fields and the BRST operator 
which played an important role in the above authors' formulations. 
In deriving BRST transformations (\ref{eq:BRST11}) 
we only compare the terms expanded in the form-degrees and 
the coefficients of quaternions 
in (\ref{eq:gecu}) and (\ref{eq:geBi}).     
The conventional ghost number for particular fields and BRST charge 
are automatically assigned by the quaternionic classifications. 

Next we can consider the physical observable. 
We can construct BRST invariant polynomials 
because of the nilpotency property of the extended differential operator. 
Bianchi identity leads to the following algebraic relation,
\begin{equation}
{\cal{Q}}{\F}^n=-[{\A}, {\F}^n].
\label{eq:ebi}
\end{equation}
Taking a trace of the gauge algebra and particular quaternion sector,
we obtain the following relations due to the vanishing nature 
of the r.h.s. of (\ref{eq:ebi}): 
\begin{eqnarray*}
 \Str_{\3}({\cal{Q}}{\F}^n) 
 &=& \Str_{\3}\Big((Q+\Q){\F}^n\Big)=0,  \\
 \Str_{\2}({\cal{Q}}{\F}^n) 
 &=& \Str_{\2}\Big((Q+\Q){\F}^n\Big)=0, 
\end{eqnarray*}
which lead to the following descendent equations:  
\begin{eqnarray}
 s\Str_{\4}({\F}^n)&=&-d\Str_{\1}({\F}^n),
\label{eq:des11} \\
 s\Str_{\1}({\cal{F}}^n)&=&-d\Str_{\4}({\F}^n).
\label{eq:des22}
\end{eqnarray}
We can then find series of gauge invariant 
physical observables:   
\begin{eqnarray}
 {\cal{O}}_{\mbox{\scr{o}}}^{\mbox{\scr{f}}}
 &=& \int_{\gamma}\Str_{\4}{\F}^n, 
\label{eq:fo}  \\
 {\cal{O}}_{\mbox{\scr{e}}}^{\mbox{\scr{b}}}
 &=& \int_{\gamma}\Str_{\1}{\F}^n,
\label{eq:be}   
\end{eqnarray}
where $\gamma$ is a homology cycle on the submanifold in 
$M$ and 
${\cal{O}}_{\mbox{\scr{o}}}^{\mbox{\scr{f}}}$ and 
${\cal{O}}_{\mbox{\scr{e}}}^{\mbox{\scr{b}}}$ denote odd-dimensional 
fermionic and even-dimensional bosonic observables, respectively.

Although we consider the above BRST algebra in two-dimensional case,
we will see that the algebra in arbitrary dimensions 
can be treated in the similar way.

We next introduce a particular model to carry out explicit analyses. 
To make the formulation concrete and simpler we specify to 
the two-dimensional flat Euclidean case 
and take the following two-dimensional antihermitian 
Euclidean Clifford algebra as the graded algebra, 
which closes under the multiplication 
and is the simplest example:  
\begin{equation}
 \begin{array}{rcl}
 T^a &:& 1, \quad \gamma_5, \\
 \Sigma^\alpha &:& \gamma^a, 
 \end{array}
\end{equation}
where $\gamma^a=(i\sigma^1, i\sigma^2)$, 
which satisfy $\{ \gamma^a, \gamma^b\}=-2\delta^{ab}$ 
and $\gamma_5=\frac{1}{2}\epsilon_{ab}\gamma^a\gamma^b
      =-i\sigma^3$ with $\epsilon_{12}=1$. 
A grading generator can be identified as $\gamma_5$ 
and then we define the supertrace  
$$
\Str(\cdots)=\Tr(\gamma_5\cdots).
$$ 

The two-dimensional topological Yang-Mills action lead 
\begin{eqnarray}
S_0&=&\frac{1}{2}\int{\mbox{Str}}_{\1}{\F}_0\wedge{\F}_0 \nonumber \\
   &=&\int d^2x\Big( \epsilon^{\mu\nu}F_{\mu\nu}|\phi|^2
                      +\epsilon^{\mu\nu}\epsilon^{ab}
                       (D_\mu\phi)_a(D_\nu\phi)_b \Big) \nonumber \\
   &=&\int d^2x\epsilon^{\mu\nu}\del_\mu\Big(2\omega_\nu|\phi|^2
                                 +\epsilon^{ab}\phi_a\del_\nu\phi_b\Big), 
\label{eq:2ty}
\end{eqnarray}
where $F_{\mu\nu}=\del_\mu\omega_\nu-\del_\nu\omega_\mu$  
and $(D_\mu\phi)_a=\del_\mu\phi_a -2{\epsilon_a}^b\omega_\mu\phi_b$.
In the action (\ref{eq:2ty}) 
the scalar part of the one-form field $\omega_{\mu s}$ 
and two-form field $B_{a\mu\nu}$ in the generalized field (\ref{eq:A0}) 
drop out because of the reducible structures of the gauge transformations. 
Then the generalized gauge transformations 
are consistently truncated to the following $SO(2)$ invariance, 
\begin{equation}
 \begin{array}{rcl}
 \delg\phi_a &=& 2v{\epsilon_a}^b\phi_b, \\
 \delg\omega_\mu &=& \del_\mu v, 
 \end{array}
\label{eq:rgt}
\end{equation}
where $v$ is a zero-form gauge parameter.
As we have discussed
we impose the topological shift symmetry, 
then BRST transformations (\ref{eq:BRST11}) lead to    
the following truncated forms: 
\begin{equation}
 \begin{array}{rcl}
 s\phi_a &=& 2{\epsilon_a}^b\phi_bC-\C_a, \\
 s\omega_\mu &=& \del_\mu C+\C_\mu, \\
 sC &=& -\eta, \\
 s\C_a &=& 2{\epsilon_a}^bC\C_b-2{\epsilon_a}^b\phi_b\eta, \\
 s\C_\mu &=& \del_\mu\eta, \\
 s\eta &=& 0,
 \end{array}
\end{equation}
where $C$, $(\C_a, \C_\mu)$, and $\eta$ are the ghost fields 
associated with $SO(2)$ gauge symmetry and topological shift symmetries, 
and the ghost for ghost field with the reducible symmetry, respectively.  
These BRST transformations indeed satisfy the nilpotency property.

We can now find out a two-dimensional instanton relation 
of our generalized gauge system by imposing the self- (anti-self-) dual 
condition 
\begin{equation}
*{\F}_0=\pm{\F}_0.
\label{eq:ic}
\end{equation}
Since a repeated application of $*$ on the generalized field strength must 
yield identity map, 
we define the following duality relation for the gauge operators  
and quaternions in addition to the usual Hodge dual operation 
on the differential forms:      
\begin{equation}
*1 = -\gamma_5, \quad 
{*}\gamma^a = -{\epsilon^a}_b\gamma^b, \quad 
{*}\gamma_5 = -1, 
\end{equation}
\begin{equation}
*{\1}={\1}, \quad 
*{\2}=-{\2}.
\end{equation}
We can then find the following minimal condition of the action 
leading to instanton relations:
\begin{eqnarray}
 & &\pm\frac{1}{2}\int\Str_{\1}{\F}_0\wedge{\F}_0
     +\frac{1}{2}\int\Str_{\1}{\F}_0\wedge*{\F}_0         \nonumber \\
 &=&\int d^2x\Bigg(\Big(\frac{1}{2}
                            \epsilon^{\mu\nu}F_{\mu\nu}\pm|\phi|^2\Big)
                            \Big(\frac{1}{2}
                            \epsilon^{\rho\sigma}F_{\rho\sigma}\pm|\phi|^2\Big)
                                                             \nonumber \\ 
 & &\qquad\quad            +\frac{1}{2}\Big((D_\mu\phi)_a
                                            \pm{\epsilon_\mu}^\nu
                                                   {\epsilon_a}^b
                                                   (D_\nu\phi)_b\Big)
                                       \Big((D^\mu\phi)^a
                                        \pm{\epsilon^\mu}_\rho
                                                   {\epsilon^a}_c
                                                   (D^\rho\phi)^c\Big)\Bigg).
\label{eq:tpi}
\end{eqnarray}
Then the instanton relations are obtained from the conditions 
for the absolute minima of the generalized Yang-Mills action
\begin{eqnarray}
 &&\frac{1}{2}\epsilon^{\mu\nu}F_{\mu\nu}-|\phi|^2 = 0,  \label{eq:sdgf1} \\
 &&(D_\mu\phi)_a^{(-)}\equiv\displaystyle{\frac{1}{2}
                           \Big((D_\mu\phi)_a-{\epsilon_\mu}^\nu
                                              {\epsilon_a}^b(D_\nu\phi)_b
                           \Big)} = 0.                           
                                                       \label{eq:sdgf2} 
\end{eqnarray}
These instanton relations are natural consequence of the formulation of the 
generalized topological Yang-Mills action as we have seen above. 

It has appeared to our attention by the recent paper~\cite{ns} 
that the dimensionally reduced 
Seiberg-Witten equation~\cite{sw} from four into two dimensions 
coincide with the eqs. (\ref{eq:sdgf1}) and (\ref{eq:sdgf2}). 
It should be noted that the Weyl spinor in Seiberg-Witten equation corresponds 
to the Higgs scalar in our formulation. 
The explicit solutions have been obtained as the Liouville vortex 
solution by Nergiz and Sa\c{c}l{\i}o\~{g}lu~\cite{ns} for the solution of 
the Seiberg-Witten equation. 
The solutions are 
%
\renewcommand{\arraystretch}{1.8}
%
\begin{equation}
  \begin{array}{rcl}
  \phi &=& \displaystyle\phi_1+i\phi_2 =\sqrt{2}\frac{dg/dz}{1-\bar{g}g}, \\ 
  \omega &=& \displaystyle\omega_\mu dx^\mu = \frac{i}{2}\Big( 
            \frac{g d\bar{g} - \bar{g} dg}{1-\bar{g}g} \Big),
  \end{array}
\label{eq:sol}
\end{equation}
%
\renewcommand{\arraystretch}{1.6}
%
where $g=g(z)$ is an arbitrary holomorphic function 
and $\bar{g}$ is the complex conjugate of $g$ with $z=x_1+ix_2$. 

The topological nature of the solutions is explicitly verified by calculating 
a flux 
\begin{equation}
\Phi = \int Fd^2x = 4\pi n, \label{eq:flux}
\end{equation}
where $F=\frac{1}{2}\epsilon_{\mu\nu}F^{\mu\nu} =|\phi|^2$. 
Here we have chosen the holomorphic function as $g(z)=z^n$. 
Due to the singular nature of the solutions (\ref{eq:sol}), 
we need particular regularization to obtain the 
explicit topological relation (\ref{eq:flux}). 
 
It is worth to mention at this stage that there is another kind of solution 
to the modified instanton relation or equivalently Bogomol'nyi equation, 
\begin{equation}
 \frac{1}{2}\epsilon^{\mu\nu}F_{\mu\nu}-|\phi|^2 + |v|^2= 0,  \label{eq:sdgf3}
\end{equation}
while the second relation (\ref{eq:sdgf2}) is the same. 
These relations yield Nielsen-Olesen vortex solution~\cite{no} which 
has again topological nature~\cite{dvf}. 
It is important to recognize that our formulation leading to the instanton 
relations (\ref{eq:sdgf1}) and (\ref{eq:sdgf2}) by the generalized topological 
Yang-Mills formulation will never lead the the kind of relation 
(\ref{eq:sdgf3}). 
Instead it may lead to the relations where $\phi$ can get constant shift: 
$\phi \rightarrow \phi + v$ which is different from (\ref{eq:sdgf3}). 
Therefore the instanton solutions obtained from the generalized topological 
Yang-Mills formulation are not Nielsen-Olesen vortex type solution but 
the dimensionally reduced one derived from four-dimensional Seiberg-Witten 
equations. 

We now derive the gauge-fixed action with instanton relations 
(\ref{eq:sdgf1}) and (\ref{eq:sdgf2})  
as gauge fixing conditions 
together with the following Landau type gauge fixing conditions  
to fix usual gauge symmetry and the reducible symmetry, 
\begin{equation}
 \del_\mu\omega^\mu = 0, \quad 
 \del_\mu\C^\mu = 0.
\label{eq:lgf1}
\end{equation}
Correspondingly we introduce a set of antighost fields 
$\lambda$, $\chi_{\mu a}$, $\etab$ and $\overline{C}$,  
and associated Lagrange multipliers,
$\widetilde{\pi}$, $\pi_{\mu a}$, $\rho$ and $\pi$. 
These fields obey the following closed BRST subalgebra,
\begin{equation}
 \begin{array}{rclcrcl}
 s\lambda&=&\widetilde{\pi}, &\qquad& s\widetilde{\pi}&=&0, \\
 s\chi_{\mu a}&=&\pi_{\mu a}, &\qquad& s\pi_{\mu a}&=&0, \\
 s\etab&=&\rho, &\qquad& s\rho&=&0, \\
 s\overline{C}&=&\pi, &\qquad&s\pi&=&0,
 \end{array}
\end{equation}
where the anti-self-dual field $\chi_{\mu a}$ 
obeys the condition 
${\epsilon_\mu}^\nu{\epsilon_a}^b\chi_{\nu b}=-\chi_{\mu a}$ 
and $\pi_{\mu a}$ also obeys the same condition.

We then obtain the following ``completely'' gauge-fixed action by 
adding the BRST-exact terms: 
\begin{eqnarray}
 S_{\mbox{\scriptsize{g-f}}}
 &=& S_0+s\int d^2x\Bigg\{
                    +\lambda\Big(\frac{1}{2}\epsilon^{\mu\nu}F_{\mu\nu}
                               -|\phi|^2-\beta\widetilde{\pi}\Big)
               -\chi_{\mu a}\Big((D^\mu\phi)^{a(-)}-\alpha\pi^{\mu a}\Big)
                                                                   \nonumber \\
 & &\qquad\qquad\qquad\quad
                +\etab\del_\mu\C^\mu+\overline{C}\del_\mu\omega^\mu\Bigg\},  
\label{eq:bs0}
\end{eqnarray}
where $\alpha$ and $\beta$ are arbitrary parameters.
Using the equations of motions for auxiliary fields 
$\pi_{\mu a}$ and 
$\widetilde{\pi}$ and choosing the parameters $\alpha=-\frac{1}{8}$ 
and $\beta=\frac{1}{4}$, we can eliminate topological sectors 
and then we obtain the physical Yang-Mills action plus fermion 
interaction terms. 

\section{Twisted $N=2$ supersymmetric action}

\setcounter{equation}{0}

We first summarize the twisting procedure of $N=2$ superalgebra. 
The algebra of $N=2$ supersymmetry without central extension is
constructed by the following relations:\footnote{
Our convention of hermite Euclidean $\gamma$-matrices is 
${(\gamma^\mu)_\alpha}^\beta=\{\sigma^1, \sigma^3\}$, where $\sigma^i$ 
are Pauli matrices, and $\gamma_5=\gamma^1\gamma^2$.
Majorana fermion is a two-dimensional real representation of $SO(2)$, 
and the Lorentz spinor indices are lowered and raised by the charge 
conjugation matrix $C_{\alpha\beta}=\delta_{\alpha\beta}$.
}
\renewcommand{\arraystretch}{1.8}
\begin{equation}
 \begin{array}{rcl}
 \{ {Q_{\alpha, i}}, Q_{\beta, j} \}
   &=&\delta_{ij}(\gamma^\mu)_{\alpha\beta}P_\mu, \\
 {[ J, P_\mu ]} &=& i{\epsilon_\mu}^\nu P_\nu, \\ 
 {[J, Q_{\alpha, i}]} &=& \displaystyle{\frac{i}{2}}
                          {(\gamma_5)_\alpha}^\beta Q_{\beta, i}, \\ 
 {[ R, Q_{\alpha, i}]} &=& {\displaystyle{\frac{i}{2}}}
                            {(\gamma_5)_i}^j Q_{\alpha, j}, \\
 {[P_\mu, Q_{\alpha, i}]} &=& {[ R, J ]} = {[ R, P_{\mu} ]} = 0.
 \end{array}
\label{eq:ssa}
\end{equation}
%
\renewcommand{\arraystretch}{1.6}
%
Here $Q_{\alpha, i}$ are the generators of supersymmetry,
where the indices $\alpha(=1, 2)$ and $i(=1, 2)$ are Lorentz spinor and
internal spinor indices labeling two different $N=2$ generators, 
respectively.  
We can take these operators to be Majorana.
$P_{\mu}$, $J$ and $R$ are generators of translation,
$SO(2)$ Lorentz rotation and internal $SO(2)_I$ rotation called
$R$-symmetry, representing spin and isospin rotation, respectively.

The above $N=2$ superalgebra is transformed into the twisted 
$N=2$ superalgebra by the following procedure.
The essential meaning of the topological twist is to identify 
the isospinor indices as spinor indices. 
Then the isospinor indices should then transform as spinor under the 
Lorentz transformation. 
This will then lead to the redefinition of the energy momentum tensor 
and the Lorentz rotation generator.
 
We consider the energy momentum tensor $T_{\mu\nu}$ and 
the conserved current $R_\mu$ associated with $R$-symmetry.  
We then define a new energy momentum tensor $T'_{\mu\nu}$ 
without spoiling the conservation law 
by the following relation: 
\begin{equation}
 T'_{\mu\nu} = T_{\mu\nu} + \epsilon_{\mu\rho}\del^\rho R_\nu +
                          \epsilon_{\nu\rho}\del^\rho R_\mu.
\end{equation}
This modification of the energy momentum tensor leads to a redefinition of 
the Lorentz rotation generator $J$, 
\begin{equation}
 J'=J+R.
\label{eq:nj}
\end{equation}   
This rotation group is interpreted as the diagonal subgroup of 
$SO(2)\times SO(2)_I$.   
Now the supercharges have double spinor indices and thus 
can be decomposed into the following scalar, pseudo scalar and vector 
components:
\begin{equation}
 {Q_\alpha}^\beta = \frac{1}{2}\Big(
                    \frac{1}{\sqrt{2}}{\delta_\alpha}^\beta \Q
                   +\sqrt{2}{(\gamma^\mu)_\alpha}^\beta Q_\mu
                   +\frac{1}{\sqrt{2}}{(\gamma_5)_\alpha}^\beta \Qt
                               \Big).
\label{eq:ttsc}
\end{equation}
Solving conversely, we obtain  
\begin{equation}
 \begin{array}{rcl}
 \Q &=& \sqrt{2}\Tr Q, \\
 \Qt &=& -\sqrt{2}\Tr(\gamma_{5}Q), \\ 
 Q_\mu &=& \displaystyle{\frac{1}{\sqrt{2}}\Tr(\gamma_{\mu}Q)}. 
 \end{array}
\end{equation}
The essence of the twisting procedure is reflected to the fact 
that the spin $1/2$ charge having the first spinor suffix turns into 
the spin $0$ or spin $1$ charge by adding the isospin $1/2$ charge, 
which can be understood by the above relations (\ref{eq:nj}) and 
(\ref{eq:ttsc}). 

The part of the algebra including Lorentz generator $J$ in (\ref{eq:ssa})
can be rewritten in terms of the new Lorentz generators $J'$ in the 
following form:
\begin{eqnarray}
 {[ J', \Q ]} &=& {[ J', \Qt ]} = 0, \nonumber \\
 {[ J', Q_{\mu} ]} &=& i{\epsilon_\mu}^\nu Q_\nu, 
\label{eq:pa} \\
 {[ J', P_{\mu} ]} &=& i{\epsilon_\mu}^\nu P_\nu, \nonumber 
\end{eqnarray}
where the scalar and vector nature of the fermionic charges 
measured by the new Lorentz generator $J'$ after the twist 
is obvious from these relations.  

The following algebra together with the algebra (\ref{eq:pa}) 
construct the twisted version of $N=2$ superalgebra, 
%
\renewcommand{\arraystretch}{2.0}
%
\begin{equation}
 \begin{array}{rclcrcl}
  \Q^2 &=& \Qt^2 = 0, &\quad&
   {[ R, \Q ]} &=& \displaystyle{\frac{i}{2}}\Qt, \\
 \{ \Q, \Qt \} &=& \{ Q_{\mu}, Q_{\nu} \} = 0, &\quad& 
   {[ R, \Qt ]} &=& -\displaystyle{\frac{i}{2}}\Q, \\ 
 \{ \Q, Q_{\mu} \} &=& 2P_{\mu}, &\quad& 
    {[ R, Q_{\mu} ]} &=& \displaystyle{\frac{i}{2}}
                      {\epsilon_\mu}^\nu Q_\nu, \\
 \{ \widetilde{Q}, Q_{\mu} \} &=& -2{\epsilon_\mu}^\nu P_\nu, &\quad& 
    {[R, J']} &=& {[R, P_\mu]} = 0. 
 \end{array}
\label{eq:ttss}
\end{equation}
%
\renewcommand{\arraystretch}{1.6}
%
Here we identify the scalar charge $\Q$ as BRST charge 
since it has a nilpotency property. 
It should be noted that the momentum operator is BRST-exact, 
which reflects the characteristic of topological field theory. 

Here we explicitly show that 
the ``partially'' gauge-fixed action possesses twisted $N=2$ 
supersymmetry~\cite{w1}. 
``Partially'' we mean to fix the gauge of topological symmetry only 
and recover the $SO(2)$ gauge symmetry.  
We first modify the gauge-fixed action (\ref{eq:bs0}) 
by adding another BRST-exact term 
$$
-2is\int d^2x\etab\epsilon_{ab}\phi^a\C^b, 
$$
and make all fields hermitian 
to assure the hermicity property of the action: 
\begin{eqnarray}
 S=\int d^2x &\Big(&+\frac{1}{2}F_{\mu\nu}F^{\mu\nu} 
                      + (D_\mu\phi)_a(D^\mu\phi)^a + |\phi|^4  \nonumber \\
             &{}& +i\rho\del_\mu\C^\mu 
                  -i\lambda\epsilon^{\mu\nu}\del_\mu\C_\nu      \nonumber \\
             &{}& -i\chi_{\mu a}(D^\mu\C)^a 
                  +\del_\mu\etab\del^\mu\eta                   \nonumber \\
             &{}& -2i\rho\epsilon^{ab}\phi_a\C_b 
                  -2i\lambda\phi^a\C_a
                  -2i\chi_{\mu a}\epsilon^{ab}\C^\mu\phi_b     \nonumber \\ 
             &{}& -\frac{i}{4}\epsilon^{\mu\nu}
                   \chi_{\mu a}{\chi_\nu}^a\eta 
                  +2i\etab\epsilon_{ab}\C^a\C^b 
                  +4\etab\eta|\phi|^2 \Big).
\label{eq:wa}
\end{eqnarray}
It is easy to see that kinetic terms of $\phi_a$, $\C_\mu$, $\rho$, 
$\lambda$, $\chi_{\mu a}$ and $\C_a$ are nondegenerate,
while that of $\omega_\mu$ is degenerate. 
Indeed this action is invariant under the following $SO(2)$ 
gauge transformations with a gauge parameter $v$:  
\begin{eqnarray}
 \delg(\phi_a, \C_a, \chi_{\mu a}) 
                 &=& 2v{\epsilon_a}^b(\phi_b, \C_b, \chi_{\mu b}), 
                     \nonumber \\
 \delg\omega_\mu &=& \del_\mu v, 
\label{eq:wgt} \\
 \delg(\C_\mu, \eta, \lambda, \rho, \etab) &=& 0. \nonumber  
\end{eqnarray}

Corresponding to the Lagrangian given in (\ref{eq:wa}), we can find explicit 
transformations of fields by the supercharges:
\begin{equation}
\theta_A s^A\varphi = {[i\theta_AQ^A, \varphi]},
\end{equation}
where $s^A=\{ s, \tilde{s}, s_\mu\}$ and $Q_A=\{\Q, \Qt, Q_\mu \}$.

We first point out that the action (\ref{eq:wa}) is invariant under 
the following BRST-like fermionic transformations: 
\begin{equation}
 \begin{array}{rclcrcl}
  s\phi_a &=& -\C_a, &\quad&   
  s\chi_{\mu a} &=& 4i(D_\mu\phi)_a^{\tiny (-)}, \\  
  s\omega_\mu &=& \C_\mu, &\quad&   
 s\lambda &=& -2i\Big(\dis\frac{1}{2}\epsilon^{\mu\nu}F_{\mu\nu} 
                      - |\phi|^2 \Big), \\
  s\C_a &=& -2i{\epsilon_a}^b\phi_b\eta, &\quad& 
  s\etab &=& \rho, \\
  s\C_{\mu} &=& i\del_\mu\eta, &\quad&
  s\rho &=& 0,   \\
  s\eta &=& 0. &&&& \\
 \end{array}
\end{equation}
These transformations are only on-shell nilpotent up to corresponding 
gauge transformations in (\ref{eq:wgt})  
\begin{equation}
 s^2=i{\delg}_{\eta},
\label{eq:ss}
\end{equation}
where ${\delg}_{\eta}$ denotes a gauge transformation 
associated with a gauge parameter $\eta$.
     
Furthermore we find that the action possesses 
the following fermionic vector symmetry: 
\begin{equation}
 \begin{array}{rclcrcl}
 s_\mu\phi_a &=& \dis{\frac{1}{2}}\chi_{\mu a}, &\quad& 
 s_\mu\chi_{\nu a} &=& -4i(\delta_{\mu\nu}\epsilon_{ab} 
                          -\epsilon_{\mu\nu}\delta_{ab})\etab\phi^b, \\
 s_\mu\omega_\nu &=& -\dis{\frac{1}{2}}
                       (\epsilon_{\mu\nu}\lambda+\delta_{\mu\nu}\rho), &\quad&
 s_\mu\lambda &=& 2i{\epsilon_\mu}^\nu\del_\nu\etab, \\
 s_\mu\C_a &=&-2i(D_\mu\phi)_a^{(+)}, &\quad& 
 s_\mu\etab &=& 0, \\ 
 s_\mu\C_\nu &=& i(F_{\mu\nu}+\epsilon_{\mu\nu}|\phi|^2), &\quad&
 s_\mu\rho &=& 2i\del_\mu\etab, \\
 s_\mu\eta &=& 2\C_\mu. &&&&\\
 \end{array}
\label{eq:sm}
\end{equation}
These transformations satisfy the following anticommutation relations:  
\begin{equation}
 \begin{array}{rcl}
 \{ s, s_\mu \} &=& 2i\del_\mu - 2i{\delg}_{\omega_{\mu}}, \\
 \{ s_\mu, s_\nu \} &=& -2i\delta_{\mu\nu}{\delg}_{\etab}, 
 \end{array}
\label{eq:ssm}
\end{equation}
where these algebras are also satisfied on shell.

Lastly we can introduce the fermionic pseudo scalar symmetry 
which are the partner of the BRST-like symmetry:  
\begin{equation}
 \begin{array}{rclcrcl}
 \st\phi_a &=& {\epsilon_a}^bs\phi_b 
                   = -{\epsilon_a}^b\C_b, &\quad&
 \st\chi_{\mu a} &=& -{\epsilon_\mu}^\nu s\chi_{\nu a}
                   = -4i{\epsilon_\mu}^\nu(D_\nu\phi)_a^{\tiny (-)}, \\
 \st\omega_\mu &=& {\epsilon_\mu}^\nu s\omega_\nu 
                   = {\epsilon_\mu}^\nu\C_\nu, &\quad&
 \st\lambda &=& 0, \\ 
 \st\C_a &=& -{\epsilon_a}^bs\C_b 
                   = -2i\phi_a\eta, &\quad& 
 \st\etab &=& -\lambda, \\ 
 \st\C_\mu &=& -{\epsilon_\mu}^\nu s\C_\nu 
                   = -i{\epsilon_\mu}^\nu\del_\nu\eta, &\quad&
 \st\rho &=& -2i\Big(\dis\frac{1}{2}\epsilon^{\mu\nu}F_{\mu\nu}
                     -|\phi|^2\Big), \\
 \st\eta &=& 0. &&&&\\
 \end{array}
\end{equation}
These transformations lead to the following anticommutation relations: 
\begin{equation}
 \begin{array}{rcl}
 \{ \st, s_\mu \} &=& -2i{\epsilon_\mu}^\nu\del_\nu 
                      + 2i{\delg}_{{\epsilon_\mu}^\nu\omega_{\nu}}, \\
 \{ \st, \st \} &=& 2i{\delg}_{\eta}, \\
  \{ \st, s \} &=& 0.
 \end{array}
\label{eq:stsm}
\end{equation}
It is apparent from (\ref{eq:ss}), (\ref{eq:ssm}) and (\ref{eq:stsm}), 
that the operators $s, s_\mu$ and $\st$ obey 
the twisted $N=2$ supersymmetric algebra.

\section{Topological twist and Dirac-K\"ahler fermion}

\setcounter{equation}{0}

In this formulation we find out very interesting correspondence.  
We point out that 
two multiplets for the ghost fields and the multiplier fields 
$(\rho, \C_\mu, \lambda)$ and $(\C_a, \chi_{\mu a})$ can be interpreted  
as Dirac-K\"ahler multiplets, as we shall see below.

In two-dimensional flat Euclidean spacetime, 
we introduce the following Dirac-K\"ahler field~\cite{g, bj} 
\begin{eqnarray}
 \Psi &=& \psi + \psi_\mu dx^\mu 
          + \frac{1}{2}\psi_{\mu\nu}dx^\mu\wedge dx^\nu \nonumber \\
      &\equiv& \sum_{\alpha, (\beta)}{\psi_\alpha}^{(\beta)}
               {Z^\alpha}_{(\beta)}, 
\end{eqnarray}
where $\psi$, $\psi_\mu$ and $\psi_{\mu\nu}$ are hermitian fermionic 
scalar, vector and antisymmetric tensor field, respectively.
The base ${Z^\alpha}_{(\beta)}$ is a $2\times 2$ 
matrix and can be expanded into the following inhomogeneous forms: 
\begin{equation}
 Z=1+\gamma_\mu^Tdx^\mu 
   + \frac{1}{2}\gamma_\mu^T\gamma_\nu^Tdx^\mu \wedge dx^\nu.
\end{equation}
The coefficient ${\psi_\alpha}^{(\beta)}$ can be 
equivalently rewritten as   
\begin{equation}
 {\psi_\alpha}^{(\beta)} = \frac{1}{2}(\psi+\psi_\mu\gamma^\mu
                         +\frac{1}{2}\epsilon^{\mu\nu}
                           \psi_{\mu\nu}\gamma_5)_\alpha^{\ \ (\beta)}.
\label{eq:bdh}
\end{equation}
It is interesting to note the remembrance of the expansion relations 
of the fermionic charge in (\ref{eq:ttsc}) and the coefficients of 
the Dirac-K\"ahler field in (\ref{eq:bdh}).
This could be understood as the origin of the Dirac-K\"ahler 
interpretation of ghost fields. 
We then find that massless Dirac equations are expressed  
as the following set of equations by the use of 
antisymmetric tensor fields: 
\begin{equation}
(d+\delta)\Psi={(\gamma^\mu\del_\mu\psi)_\alpha}^{(\beta)}
                {Z^\alpha}_{(\beta)}=0,
\end{equation}
where $\delta$ is an adjoint operator $\delta=*d*$ 
and the index $\alpha$ is a spinor one while 
the index $(\beta)$ is regarded as a ``flavour'' one for two degenerate 
Dirac fermions. 
The Dirac-K\"ahler action which leads to the above equation of motion 
is defined by 
\begin{eqnarray}
S&=&\frac{1}{2}\int i\Psi^* \wedge * (d+\delta)\Psi \nonumber \\
 &=&\int d^2x\sum_{(\beta)}i{(\psi^{\dagger})_{(\beta)}}^\alpha
                           {(\gamma^\mu\del_\mu\psi)_{\alpha}}^{(\beta)}
                                                  \nonumber \\
 &=&\int d^2x \Tr\Big(i\psi^{\dagger}\gamma^\mu\del_\mu\psi\Big).
\end{eqnarray}

We now turn to describe ghost fields in terms of Dirac-K\"ahler fields.   
The kinetic terms of these multiplets in the action (\ref{eq:wa}) 
can be expressed as
\begin{eqnarray}
    && \int d^2x(i\rho\del_\mu\C^\mu-i\lambda\epsilon^{\mu\nu}\del_\mu\C_\nu
                -i\chi^{\mu a}\del_\mu\C_a)                      \nonumber \\
   &=& \int d^2x\Tr\Big(i\psi^\dagger\gamma^\mu\del_\mu\psi
                        +i\chi^\dagger\gamma^\mu\del_\mu\chi\Big),
\end{eqnarray}
where Dirac-K\"ahler fields $\psi$ and $\chi$ are given by
\begin{equation}
 \begin{array}{rcl}
 \psi &=& \frac{1}{2}(\rho+\C_\mu\gamma^\mu-\lambda\gamma_5), \\
 \chi &=& \frac{1}{2}(-\C_{a=1}+\chi_{\mu a=1}\gamma^\mu-\C_{a=2}\gamma_5). \\
 \end{array}
\label{eq:dkp}
\end{equation}
Here we impose anti-self-dual conditions for $\chi_{\mu a}$. 
It is easy to see that the action possess 
$SO(2)$ ``flavour'' symmetry.  

The final expression of the twisted $N=2$ supersymmetric action 
with Dirac-K\"ahler fermions is 
\begin{eqnarray}
 S=\int d^2x\Big(&+&\frac{1}{2}F_{\mu\nu}F^{\mu\nu} 
                 + (D_\mu\phi)_a(D^\mu\phi)^a + |\phi|^4  \nonumber \\
             &+&\frac{1}{2}\del_\mu A\del^\mu A 
                -\frac{1}{2}\del_\mu B\del^\mu B \nonumber \\
             &+& \Tr(i\psi^\dagger\gamma^\mu\del_\mu\psi)
                +\Tr(i\chi^\dagger\gamma^\mu D_\mu\chi)   \nonumber \\
             &-& 4i\phi_1\Tr(\psi^\dagger\gamma_5\chi)
                 +4i\phi_2\Tr(\psi^\dagger\gamma_5\chi\gamma_5)  \nonumber \\
             &-& i\sqrt{2}A\Tr(\chi^\dagger\gamma_5\chi)
               +i\sqrt{2}B\Tr(\chi^\dagger\chi\gamma_5) \nonumber \\
             &+& 2(A^2-B^2)|\phi|^2\Big), 
\label{eq:dk}
\end{eqnarray}
where we denote $\eta\equiv\frac{2}{\sqrt{2}}(A+B)$ and    
$\etab\equiv\frac{1}{2\sqrt{2}}(A-B)$ 
and the covariant derivative with respect to the flavour group on 
Dirac-K\"ahler field $\chi$ is given by 
$D_\mu\chi\equiv\del_\mu\chi+2\omega_\mu\chi\gamma_5$.
It is worth to mention that this action 
is equivalent to the extended supersymmetric abelin Higgs system~\cite{dvf} 
and topological Bogomol'nyi theory~\cite{st}
except for the symmetry breaking potential. 

As we have seen in the formulation, the fermionic fields appearing 
in the quantization procedure such as ghost fields 
turns into the Dirac-K\"ahler matter fermion. 
It would be important to confirm algebraically 
that the Dirac-K\"ahler fermions tranform as spinor fields 
and possess half-integral spin unlike the ghost fields.  

Redefining the Lorentz generator, we will perform a change of the spin 
of the operators. 
Indeed we will assign $R$-quantum number integral and half-integral 
for boson fields and fermion fields, respectively.    
The twisted $N=2$ theory defined by $J'$  
is the topological field theory, 
whose superalgebra corresponds to 
(\ref{eq:pa}) and (\ref{eq:ttss}), 
while the theory defined by $J$ is $N=2$ supersymmetric field theory.   

It is important to recognize that in the present model 
we can identify the $R$-symmetry as the flavour symmetry of the 
Dirac-K\"ahler fields 
\begin{equation}
 \delta_{R}\psi = \psi\Big(\frac{1}{2}\gamma_5\Big), \qquad 
 \delta_{R}\chi = \chi\Big(\frac{1}{2}\gamma_5\Big), 
\end{equation}
which should be compatible with the algebra (\ref{eq:ttss}).
The origin of this identification is again due to the resemblance of 
the relations between (\ref{eq:ttsc}) and (\ref{eq:bdh}). 
In other words this identification is originated from the 
observation that the second flavour suffix of the Dirac-K\"ahler field 
in eq. (\ref{eq:bdh}) has faithful correspondence with the second 
spinor suffix of the fermionic charge in eq. (\ref{eq:ttsc}), 
which originally corresponds to the isospinor suffix of $R$-generator 
before the twist. 
Then Lorentz transformation on the Dirac-K\"ahler field $\psi$ 
induced by $J'$ is 
\begin{eqnarray}
 \delta_{J'}\psi &=& \frac{1}{2}
                     \Big(-{\epsilon_\mu}^\nu\C_\nu\gamma^\mu\Big) \nonumber \\
                 &=& -\frac{1}{2}{[\gamma_5, \psi]}.
\end{eqnarray}
On the other hand the Lorentz transformation induced by $J=J'-R$ is 
\begin{eqnarray}
 \delta_J\psi &=& \delta_{J'}\psi - \delta_R\psi \nonumber \\
              &=& -\frac{1}{2}\gamma_5\psi, 
\end{eqnarray}
which precisely coincides with the Lorentz transformation of spinor field. 
This implies that Dirac-K\"ahler field is exactly transformed as
spinors.  
We can obtain the same relation for $\chi$. 
Consequently we have found that the twisting mechanism in
two-dimensional $N=2$ theory has been understood 
from the Dirac-K\"ahler fermion formulation and the $R$-symmetry 
is nothing but the flavour symmetry of the Dirac-K\"ahler fermion. 

\section{Conclusions and discussions}

We have investigated the generalized gauge theory 
from the topological field theory point of view. 
Firstly we have extended the algebraic structure of BRST 
transformation \'a la Baulieu-Singer and derived sets of BRST 
invariant physical operators.  
This extension fits naturally in the framework 
of the generalized gauge theory.
The classical fields, ghost fields, ghost for ghost fields, the 
differential operator and BRST operator are treated 
in a unified way by the quaternion algebra. 
In particular commutator and anti-commutator difference in the 
algebra is automatically accomodated in the generalized gauge 
theory formulation while this point is treated in an ad hoc 
way in the previous treatments.

As a concrete example
we have quantized the generalized topological Yang-Mills action 
in two-dimensional flat Euclidean spacetime 
with the two-dimensional Clifford algebra as the simplest graded Lie 
algebra.
We have shown that the generalized topological Yang-Mills action 
is BRST equivalent to the standard Yang-Mills action plus fermionic 
ghost and Lagrange multiplier terms by imposing the instanton relations 
as the gauge fixing conditions. 
It turns out that the instanton relations coincide with 
the two-dimensional counterpart of the Seiberg-Witten relations 
dimensionally reduced from four into two dimensions.
The explicit topological solutions of the instanton relations have 
been obtained~\cite{ns}. 
The full twisted $N=2$ supersymmetric algebra has been examined 
for the gauge-fixed action and explicit transformations of fields 
for the fermionic charge family included BRST charge has 
been obtained.  

We found that the fermionic sector including ghost fields in the 
gauge-fixed action can be identified with the Dirac-K\"ahler 
fermions. 
The crucial observation is that the $R$-symmetry of $N=2$ 
supersymmetric action can be identified with the ``flavour'' symmetry 
of the Dirac-K\"ahler fermion action.
Then the ghost fields together with the fermionic multiplier 
fields turn into matter fermions via twisting mechanism.
On the other hand the twisting mechanism is equivalent to the 
Dirac-K\"ahler fermion formulation when we identify the 
$R$-symmetry and the ``flavour'' symmetry. 
In this sense we have found that the twisting mechanism is essentially 
equivalent to the generation of matter fermions from fermionic 
ghosts via Dirac-K\"ahler fermion formulation. 
It is interesting to see if this mechanism works even in higher 
dimensions.
    
\vspace{1cm}

\noindent
{\Large{\bf Acknowledgments}}

This work is supported in part by Japan Society for the 
Promotion of Science under the grant number 09640330. 


\end{document}